\documentclass[conference]{IEEEtran}
\IEEEoverridecommandlockouts
\usepackage{cite}
\usepackage{amsmath,amssymb,amsfonts}
\usepackage{algorithmic}
\usepackage{graphicx}
\usepackage{textcomp}
\usepackage{xcolor}
\usepackage{booktabs}
\usepackage[hidelinks]{hyperref}

\def\BibTeX{{\rm B\kern-.05em{\sc i\kern-.025em b}\kern-.08em
    T\kern-.1667em\lower.7ex\hbox{E}\kern-.125emX}}
\begin{document}

\title{Development of Different Algorithms for Drone-Based Antenna Measurement Systems and Near-Field Error Analysis}

\author{
\begin{tabular}{ccc}
\begin{minipage}[t]{0.32\textwidth}
\centering
Simranjit Singh\\
\textit{Indian Institute of Technology Delhi}\\
New Delhi, India\\
eee252834@iitd.ac.in
\end{minipage}
&
\begin{minipage}[t]{0.30\textwidth}
\centering
Jaswant Sharma\\
\textit{Space Applications Centre, ISRO}\\
Ahmedabad, India\\
jaswantamd@sac.isro.gov.in
\end{minipage}
&
\begin{minipage}[t]{0.30\textwidth}
\centering
Jigar M.~Pandya\\
\textit{Space Applications Centre, ISRO}\\
Ahmedabad, India\\
jigar@sac.isro.gov.in
\end{minipage}
\end{tabular}
\vspace{-1em}
}

\maketitle

\begin{abstract}
Near-field antenna measurements underpin the characterization of electrically large apertures, yet the fidelity of the Near-Field to Far-Field (NF-FF) transformation depends on the reconstruction algorithm's assumptions and robustness to real-world imperfections, including those from drone-based scanning platforms.

Classical FFT-based modal expansion is efficient on uniformly sampled canonical grids but fails when phase-coherent acquisition cannot be maintained. We address this via a phaseless NF-FF algorithm reconstructing the far field from amplitude-only data through iterative phase retrieval. When sampling becomes sparse or irregular, even amplitude-based methods break down, motivating the \textbf{Adaptive Sparse Inverse Radiation Estimator (ASPIRE)}, a full-complex inverse source framework that solves a Method-of-Moments problem over RWG basis functions via cascaded rSVD and regularized shrinkage. Mutual coupling between basis functions is explicitly resolved, improving reconstruction fidelity beyond coupling-agnostic inverse-source formulations. The solver is accelerated via a Multilevel Fast Multipole Method engine with Numba just-in-time compilation, achieving a $1.2\times$ reduction in matrix-vector product time and up to $15\times$ lower memory usage relative to dense evaluation at N=100K.

Across frequency bands and positioning/truncation error scenarios, the pipeline sustains algorithmic stability and achieves sub-degree beamwidth reconstruction error. These results establish an error-aware framework for algorithm selection across fixed and drone-based near-field measurement platforms.
\end{abstract}


\section{Introduction}

Accurate characterization of antenna radiation patterns is essential for communication and radar system deployment~\cite{b2}. For electrically large apertures, direct far-field measurement demands prohibitively large separation distances, making near-field (NF) techniques the standard approach. Conventional NF-FF transformation relies on planar, cylindrical, or spherical scanning inside anechoic chambers with precision positioners~\cite{b2}. Such facilities are expensive, and the antenna size they can accommodate is physically constrained.

Drone-based antenna measurement systems (DAMS) have emerged as a cost-effective alternative for outdoor near-field scanning~\cite{b1}. By mounting a probe on a UAV equipped with RTK-GNSS positioning, NF data can be acquired over flexible scan surfaces without chamber infrastructure. However, drone platforms introduce error sources absent in chamber measurements: (i)~positioning jitter from wind loading and flight-controller dynamics, (ii)~scan-area truncation imposed by battery endurance, and (iii)~potential loss of phase coherence between transmit reference and receive chains over extended flight durations.

The classical FFT-modal approach~\cite{b4,b5} assumes uniform sampling on a canonical planar grid with full phase information. Under drone-induced positioning errors on the order of several millimeters, spectral leakage and aliasing degrade sidelobe reconstruction. Phaseless NF-FF methods based on Gerchberg--Saxton iteration~\cite{b6,b7} relax the phase-coherence requirement by operating on amplitude-only data from two measurement planes but remain confined to uniform grids and are sensitive to plane separation. Inverse-source methods~\cite{b11,b12} solve for equivalent currents on a Huygens surface enclosing the antenna under test (AUT), offering a more general framework. However, prior formulations typically employ simplified coupling models, limiting fidelity at higher frequencies where mutual coupling between basis functions is significant.

This paper presents the following contributions:
\begin{enumerate}
    \item ASPIRE, a full-complex inverse-source solver using RWG basis functions with explicit mutual coupling;
    \item A cascaded randomized SVD and $\ell_1$-regularized iterative shrinkage solver (ISS) with automatic rank and regularization parameter selection;
    \item MLFMM acceleration with Numba JIT compilation, reducing memory by up to $15\times$ and enabling scaling to $10^5$+ unknowns;
    \item A comparative error analysis across FFT-modal, phaseless GS, and ASPIRE, demonstrating ASPIRE's $\leq 0.3^\circ$ beamwidth error at C-band and X-band.
\end{enumerate}

\section{Related Work and Distinct Contributions}

Classical NF-FF transformation via FFT-based modal expansion~\cite{b4,b5} requires uniform sampling and full phase coherence-conditions frequently violated by drone platforms. While Newell~\cite{b3} established standard error analysis for planar scanning, it assumes precision positioners rather than stochastic UAV jitter. Phaseless approaches based on Gerchberg--Saxton iteration~\cite{b19,b6} and its antenna-specific variants~\cite{b7,b8,b9} successfully relax phase-hardware requirements by recovering phase from amplitude-only measurements on two planes. However, unlike our proposed approach, these methods remain strictly bound to uniform canonical grids and suffer severe degradation under scan-area truncation. 
Inverse-source methods reconstruct equivalent currents on a closed surface enclosing the AUT~\cite{b11,b12}. While Schnattinger et al.~\cite{b10} successfully combined inverse-source reconstruction with MLFMM acceleration for phaseless data, their formulation does not explicitly address the physical inter-element mutual coupling critical at higher frequencies, nor the unique truncation penalties of drone platforms. 
In our prior work~\cite{b1}, we introduced the physical drone measurement platform (DAMS) and demonstrated preliminary C-band results using an unaccelerated ASPIRE solver. The present paper distinctly shifts the contribution from mechanical positioning to algorithmic benchmarking and computational scaling. Specifically, this work provides a rigorous comparative error analysis across FFT-modal, phaseless GS, and ASPIRE under X-band UAV jitter conditions; integrates MLFMM with Numba JIT compilation for $\mathcal{O}(N\log N)$ scaling; and establishes a formal algorithm selection framework for irregular near-field acquisitions.
\section{Problem Formulation and Error Model}

Consider a planar near-field scan at standoff distance $z_0$ from the AUT aperture. A drone-mounted probe traverses an $L \times L$ grid of $M$ spatial points, recording the complex voltage at each position~$\mathbf{r}_m$. In the inverse-source formulation, the AUT is enclosed by a Huygens surface $\mathcal{S}$ discretized into triangular facets carrying RWG basis functions~\cite{b17}. The forward model relating the $N$ unknown equivalent-current coefficients $\mathbf{x} \in \mathbb{C}^N$ to the measured voltages $\mathbf{u} \in \mathbb{C}^M$ is
\begin{equation}
    \mathbf{u} = \mathbf{G}\,\mathbf{x} + \mathbf{n},
    \label{eq:forward}
\end{equation}
where $\mathbf{G} \in \mathbb{C}^{M \times N}$ is the forward operator encoding the free-space dyadic Green's function evaluated between each RWG basis function and each probe location, and~$\mathbf{n}$ accounts for measurement noise and model mismatch. Because $\mathbf{G}$ is constructed from the full MoM impedance formulation, inter-element mutual coupling is inherently captured.

Three error sources dominate drone-based NF measurements.
\textit{Positioning jitter}: deviations of the probe from its nominal grid position, characterized by lateral standard deviation $\sigma_{xy}$ and axial standard deviation $\sigma_z$. At X-band ($\lambda = 31.56$~mm), typical drone jitter of $\sigma_{xy} \approx 3.8$~mm and $\sigma_z \approx 5.75$~mm corresponds to $0.12\lambda$ and $0.18\lambda$, respectively.
\textit{Truncation}: the finite scan aperture excludes wide-angle radiation, producing Gibbs-type oscillations in the reconstructed pattern.
\textit{Phase coherence}: over extended acquisition times, the transmit--receive phase reference may drift, motivating phaseless formulations that operate on $|\mathbf{u}|$ alone.

\section{Proposed Method}

\subsection{Baseline Methods}

\textit{FFT-Modal.}
The classical NF-FF transformation computes the plane-wave spectrum via 2-D FFT of the uniformly sampled near-field data, applies a phase-correction propagator to remove the measurement-plane phase tilt, and evaluates the far-field pattern at the desired observation angles~\cite{b5}. The method is computationally efficient ($\mathcal{O}(M\log M)$) but requires uniform sampling, full phase data, and sufficient scan extent.

\textit{Phaseless GS.}
The Gerchberg--Saxton algorithm~\cite{b19} recovers the near-field phase from amplitude measurements on two parallel planes separated by $\Delta z$~\cite{b6}. The field is iteratively propagated between planes via angular-spectrum propagation; at each plane, the measured amplitude is enforced while the phase evolves freely. Convergence depends on the plane separation ($\Delta z = 3\lambda$ is used here) and on the amplitude diversity between planes~\cite{b18}.

\subsection{ASPIRE Core}

ASPIRE solves the underdetermined inverse problem~\eqref{eq:forward} by exploiting the physical sparsity of equivalent currents-only conducting elements of the AUT carry significant current, while substrate and surrounding facets do not. The full pipeline is detailed in our prior work~\cite{b1}; we summarize its two stages below.

\textit{Stage~1-Dimensionality reduction.}
A randomized SVD~\cite{b14} of $\mathbf{G}$ identifies the numerical rank~$k$ via singular-value decay against the noise floor, projecting $\mathbb{C}^N\!\to\!\mathbb{C}^k$ and suppressing noise amplification.

\textit{Stage~2-Sparse recovery.}
The reduced problem is solved via Iterative Shrinkage-Solver (ISS)~\cite{b13}:
\begin{equation}
    \mathbf{z}_t = \mathbf{x}_t - \frac{1}{L_{\mathrm{lip}}} \mathbf{G}^H\!\left(\mathbf{G}\mathbf{x}_t - \mathbf{u}\right),
    \label{eq:iss_grad}
\end{equation}
\begin{equation}
    x_{t+1}^{(i)} = \frac{z_t^{(i)}}{\bigl|z_t^{(i)}\bigr|}\,\max\!\left(\bigl|z_t^{(i)}\bigr| - \frac{\lambda}{L_{\mathrm{lip}}},\; 0\right),
    \label{eq:iss_thresh}
\end{equation}
where $L_{\mathrm{lip}}$ is the Lipschitz constant of $\mathbf{G}^H\mathbf{G}$ and $\lambda$ is chosen empirically; \eqref{eq:iss_thresh} zeros coefficients below $\lambda/L_{\mathrm{lip}}$, nullifying current on non-radiating elements.

At C-band, FISTA acceleration~\cite{b13} with Nesterov momentum provides faster convergence. At X-band, the shorter wavelength increases the condition number of $\mathbf{G}$, causing FISTA's momentum term to overshoot and destabilize. The memoryless ISS variant~\eqref{eq:iss_grad}--\eqref{eq:iss_thresh} provides robust convergence in this regime.

The recovered $\hat{\mathbf{x}}$ is debiased via least-squares on its support $\mathcal{S}^* = \{i : |\hat{x}_i| > 0\}$ before far-field synthesis, $\mathbf{E}_{\mathrm{FF}} = \mathbf{F}\hat{\mathbf{x}}$, $\mathbf{F} \in \mathbb{C}^{P \times N}$.

\subsection{MLFMM Acceleration with JIT Compilation}

The X-band forward operator ($11{,}025 \times 45{,}000$, $\approx 7.94$~GB) requires one forward and one adjoint MVP per ISS iteration, making dense evaluation memory-prohibitive at larger $N$.

We instead compute MVPs via a Multilevel Fast Multipole Method~\cite{b15,b16} in $\mathcal{O}(N\log N)$ time and memory, replacing $\mathcal{O}(N^2)$ dense operations. The kernel is implemented in pure Python with Numba JIT compilation, avoiding Fortran/C dependencies while achieving near-native speed.


\section{Results and Discussion}

\subsection{Experimental Setup}

Two AUT systems are used for validation: a C-band planar array at 6.7125~GHz~\cite{b1} and an X-band planar array at 9.5~GHz. Table~\ref{tab:xband_specs} summarizes the X-band system specifications. Reference truth patterns are obtained from Compact Antenna Test Range (CATR) measurements.

\begin{table}[htbp]
\centering
\caption{System Specifications and UAV Flight Parameters for the X-Band Testing Campaign}
\label{tab:xband_specs}
\begin{tabular}{@{}ll@{}}
\toprule
\textbf{Parameter} & \textbf{Value} \\
\midrule
\multicolumn{2}{c}{\textbf{AUT, Probe, and Huygens Surface}} \\
AUT & X-band array, $370 \times 370$ mm aperture \\
Frequency & 9.5 GHz (Wavelength, $\lambda = 31.56$ mm) \\
Probe & WR90 OEWG \\
Huygens surface, $S$ & $401.6 \times 401.6 \times 31.6$ mm box \\
Surface standoff & $\lambda/2$ from AUT aperture \\
Mesh & 30,001 triangles, 15,003 vertices \\
RWG basis functions, $N$ & \textbf{45,000} \\
\midrule
\multicolumn{2}{c}{\textbf{Near-Field Measurement Grid}} \\
Plane standoff, $z$ & 150 mm \\
Grid (extent, spacing) & $\pm 625$ mm, $\approx 12.02$ mm ($105 \times 105$) \\
Spatial points $\rightarrow M$ & \textbf{M = 11,025} (Single Polarization) \\
Drone positioning jitter & $\sigma_{xy} = 3.8$ mm and $\sigma_z = 5.75$ mm \\
\midrule
\multicolumn{2}{c}{\textbf{System Matrices Memory Footprint}} \\
Voltage vector, $\mathbf{u} \in \mathbb{C}^{M}$ & 11,025 elements ($\approx 176$ KB) \\
Forward Operator, $\mathbf{G} \in \mathbb{C}^{M \times N}$ & $11,025 \times 45,000$ ($\approx 7.94$ GB) \\
Far-Field Map, $\mathbf{F} \in \mathbb{C}^{8,406 \times N}$ & $8,406 \times 45,000$ ($\approx 6.05$ GB) \\
\bottomrule
\end{tabular}
\end{table}

\subsection{Far-Field Reconstruction Accuracy}

Table~\ref{tab:comparison} summarizes $\Delta_{\mathrm{BW}}$ across both bands and runtime for X-band. ASPIRE attains the lowest error at both frequencies ($-0.30^\circ$ C-band, $-0.20^\circ$ X-band). The smaller absolute baseline errors at X-band reflect the array's higher directivity (true BW $=4.4^\circ$), not improved robustness: relative to true BW, FFT-modal error rises from $5.9\%$ to $6.8\%$, consistent with jitter forming a larger fraction of $\lambda$ ($0.08\lambda \to 0.12\lambda$). Phaseless GS shows the opposite trend ($14.1\%\to6.8\%$), which isolates scan-truncation-not jitter-as the dominant error source at C-band, since amplitude-only retrieval is disproportionately sensitive to abruptly clipped wide-angle energy.


\begin{table}[t]
\caption{Comparison of NF-FF Methods}
\label{tab:comparison}
\centering
\begin{tabular}{lccccc}
\toprule
\textbf{Method} & \textbf{Phase} & \textbf{Uniform} & \textbf{C-Band} & \textbf{X-Band} & \textbf{Run-} \\
                & \textbf{Req.}  & \textbf{Grid}    & $\Delta_\text{BW}$ & $\Delta_\text{BW}$ & \textbf{time} \\
\midrule
FFT-Modal    & Yes & Yes & $-1.10^\circ$ & $+0.30^\circ$ & $<1$~s \\
Phaseless GS & No  & Yes & $-2.60^\circ$ & $-0.30^\circ$ & ${\sim}60$~s \\
\textbf{ASPIRE} & Yes & \textbf{No} & $\mathbf{-0.30^\circ}$ & $\mathbf{-0.20^\circ}$ & ${\sim}15.2$~min \\

\bottomrule
\end{tabular}

\raggedright \footnotesize \textit \\ 
{Note:} Runtimes evaluated on an Apple Silicon M-Series processor.
\end{table}

\begin{figure}[t]
\centering
\includegraphics[width=\columnwidth]{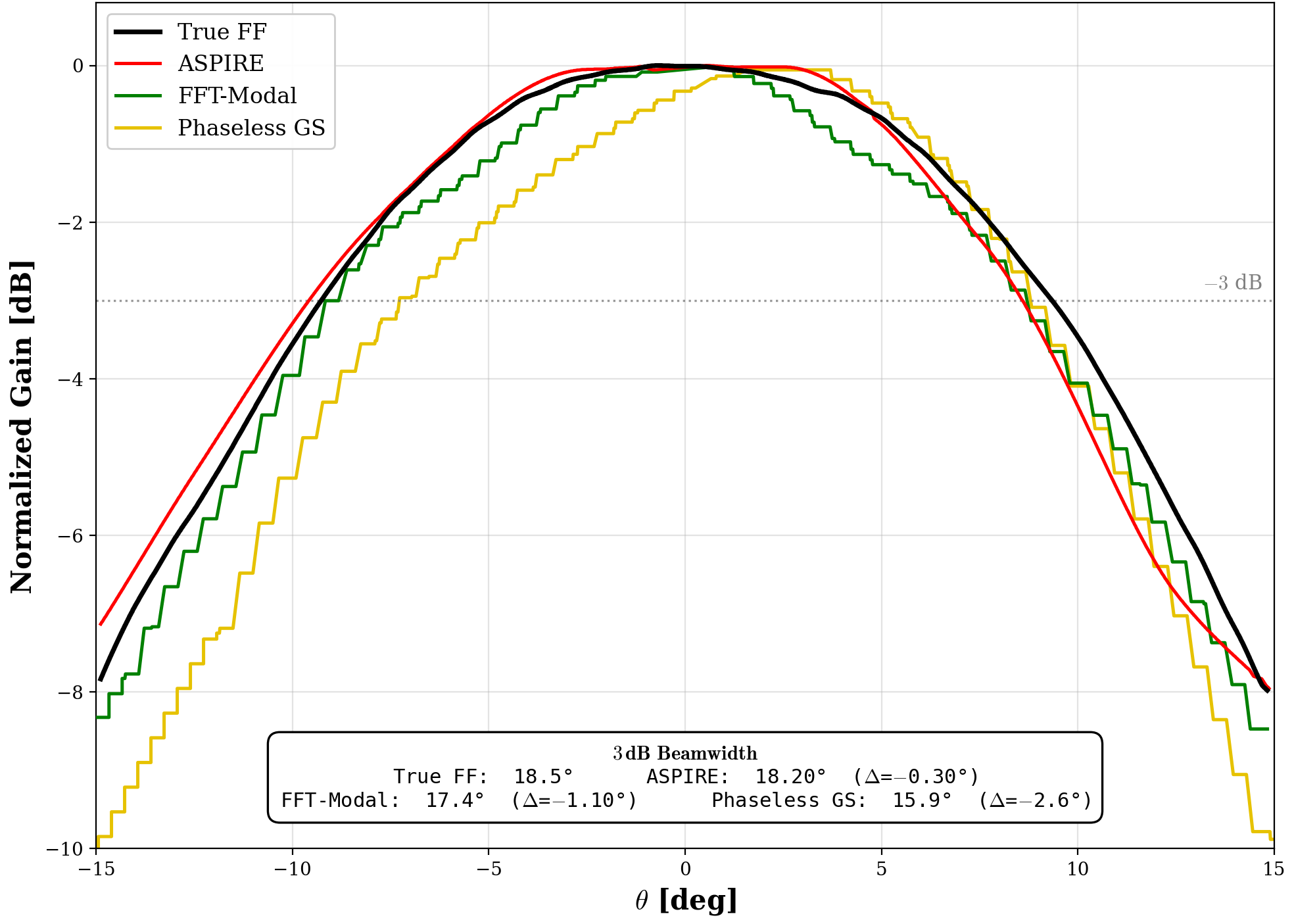}
\caption{C-band ($f = 6.7125$~GHz) main beam pattern comparison, $\varphi = 0^\circ$ (E-plane). The 3~dB beamwidths and deviations from CATR truth are annotated. ASPIRE reconstruction is performed with $0.1^\circ$ resolution, whereas the prior work~\cite{b1} used a $5^\circ$ resolution.}
\label{fig:cband}
\end{figure}

Fig.~\ref{fig:cband} shows this truncation effect at C-band. ASPIRE tracks CATR truth closely ($3$~dB BW: $18.20^\circ$ vs.\ $18.50^\circ$), while FFT-modal ($17.4^\circ$) and GS ($15.9^\circ$) narrow sharply. The array's broad $18.5^\circ$ beam places significant energy at the drone grid's edge; GS's plane-to-plane angular-spectrum propagation is especially sensitive to this boundary, driving its $14.1\%$ error. ASPIRE's $\ell_1$-regularized inverse-source formulation instead confines radiation to the Huygens surface, effectively extrapolating the missing wide-angle data and avoiding truncation-induced narrowing.

\begin{figure}[t]
\centering
\includegraphics[width=\columnwidth]{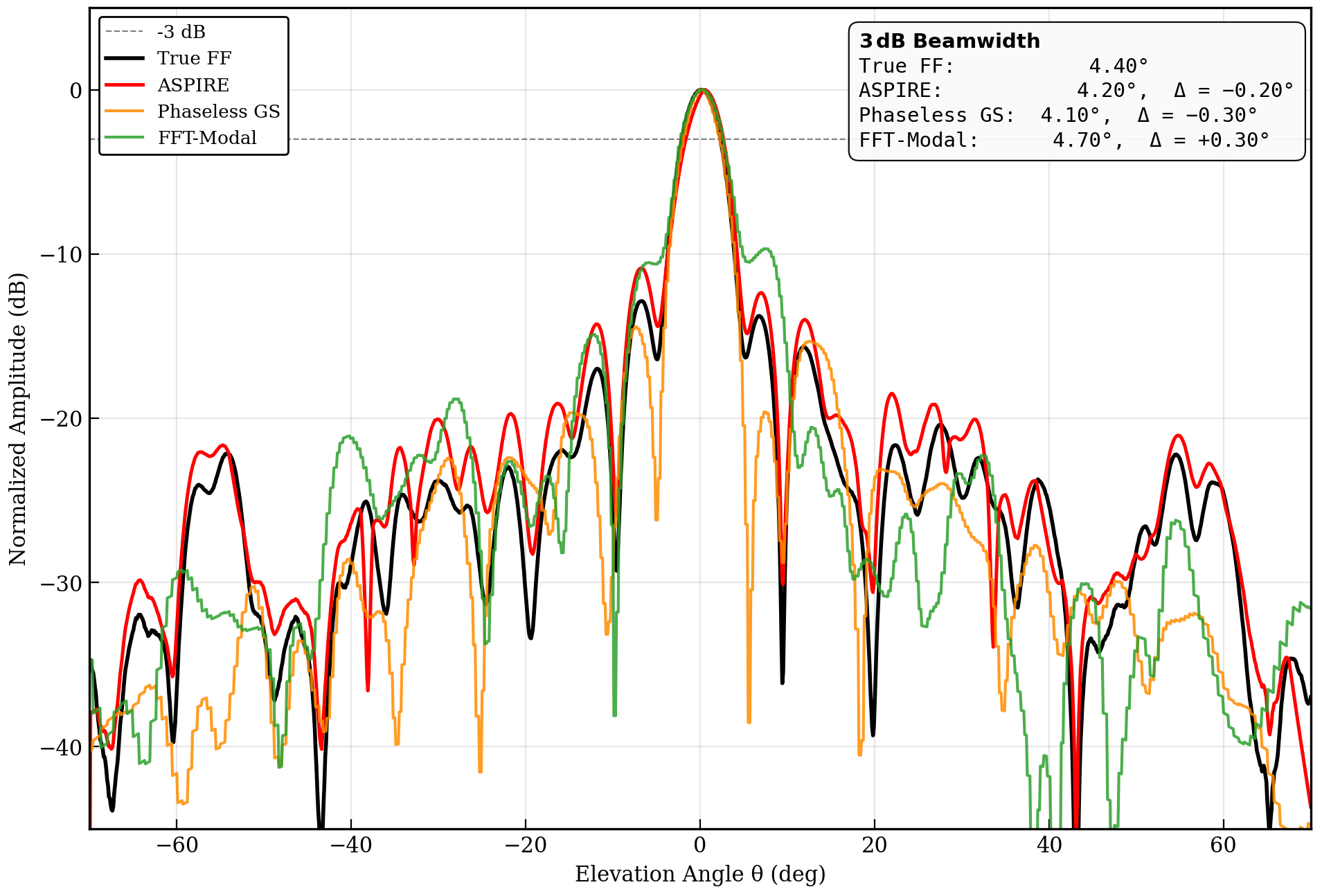}
\caption{X-band ($f = 9.5$~GHz) far-field pattern, $\varphi = 0^\circ$ (E-plane). ASPIRE (red) tracks the CATR truth (black) across the full $\pm 70^\circ$ range.}
\label{fig:xband}
\end{figure}

Fig.~\ref{fig:xband} shows the X-band pattern over $\pm70^\circ$. ASPIRE best matches CATR truth (Pearson $r=0.97$ in dB), versus FFT-modal ($r=0.87$) and GS ($r=0.82$). The gap is largest beyond $\pm20^\circ$, where the RWG forward model regularizes the high-frequency spectral content that jitter otherwise amplifies in FFT-modal processing.

\subsection{Computational Scaling}

\begin{figure}[t]
\centering
\includegraphics[width=\columnwidth]{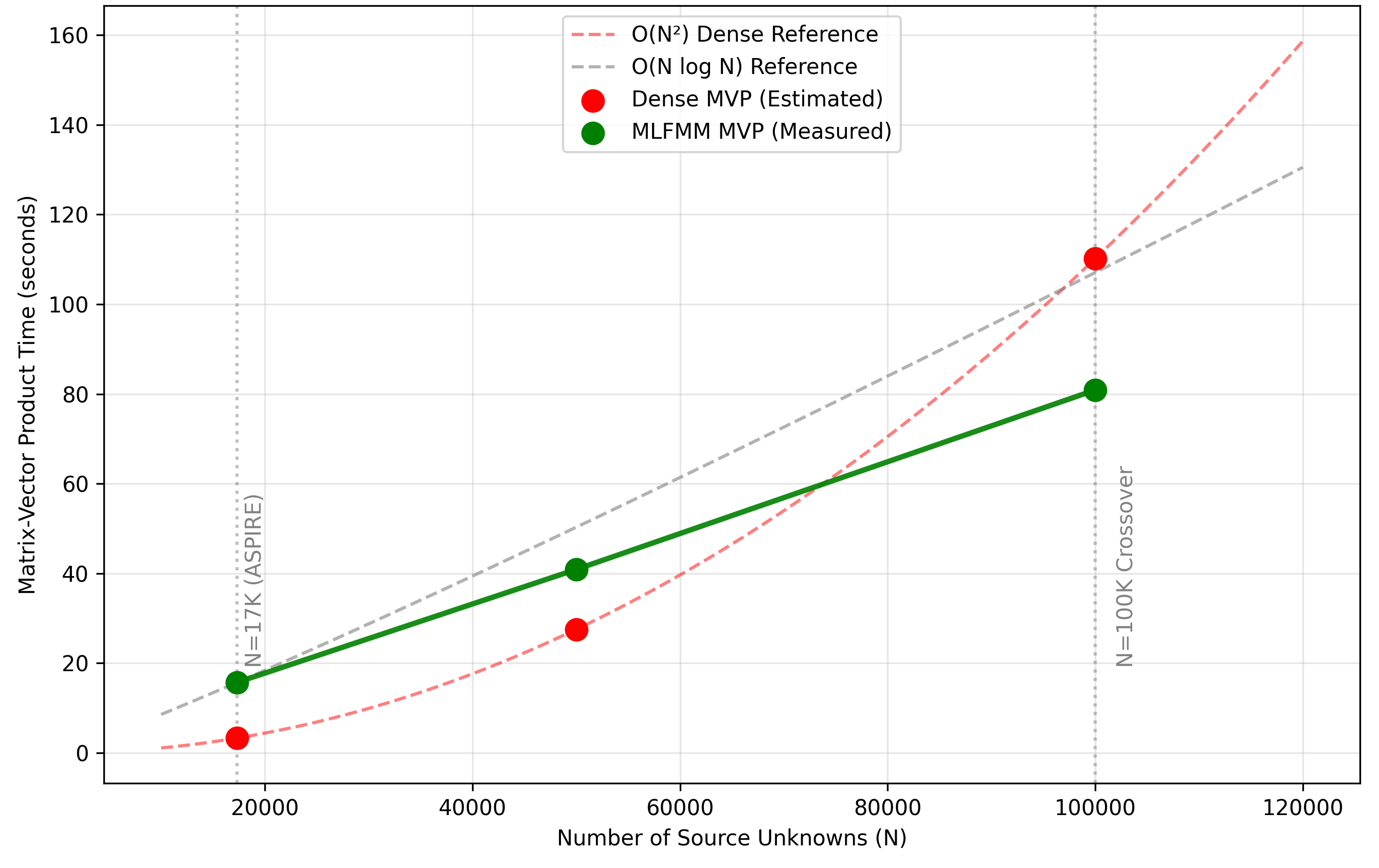}
\caption{Matrix-vector product time vs.\ number of source unknowns $N$. MLFMM (Numba JIT) scales as $\mathcal{O}(N\log N)$, crossing below the $\mathcal{O}(N^2)$ dense curve near $N = 100$K. The large-scale dense performance and storage numbers are estimated assuming $\mathcal{O}(N^2)$ complexity.}
\label{fig:mlfmm}
\end{figure}

Fig.~\ref{fig:mlfmm} shows the MVP execution time as a function of source unknowns~$N$ for both dense and MLFMM evaluation. The crossover point occurs near $N \approx 100$K, beyond which MLFMM provides both speed and memory advantages. Table~\ref{tab:mlfmm_scaling} details the performance scaling benchmark. For the specific X-band configuration evaluated in this study ($N = 45$K), the full ASPIRE pipeline required 15.2~minutes to run using the MLFMM engine. At $N = 100$K, dense storage requires 48~GB of RAM, while MLFMM operates within 3.06~GB-a $15.7\times$ reduction that makes the solver practical on commodity hardware.

\begin{table}[htbp]
\centering
\caption{Performance Scaling Benchmark: Dense vs. MLFMM}
\label{tab:mlfmm_scaling}
\resizebox{\columnwidth}{!}{
\begin{tabular}{@{}l c c c c c c@{}}
\toprule
\textbf{Scale ($N \times M$)} & \multicolumn{2}{c}{\textbf{17K $\times$ 6K}} & \multicolumn{2}{c}{\textbf{45K $\times$ 11K}} & \multicolumn{2}{c}{\textbf{100K $\times$ 30K}} \\
\cmidrule(lr){2-3} \cmidrule(lr){4-5} \cmidrule(lr){6-7}
\textbf{Metric} & Dense & FMM & Dense* & FMM & Dense* & FMM \\
\midrule
Setup Time (s) & $\sim 0$ & 20.0 & $\sim 0$ & 50.0 & $\sim 0$ & 113.0 \\
Forward MVP (s) & 3.3 & 15.7 & 15.7 & 37.1 & 94.6 & 80.9 \\
Adjoint MVP (s) & 3.3 & 12.1 & 15.7 & 22.5 & 94.6 & 73.0 \\
RAM Usage (GB) & 1.67 & \textbf{0.2} & 7.95 & \textbf{1.05} & 48.0 & \textbf{3.06} \\
\bottomrule
\end{tabular}
}
\end{table}

\subsection{Algorithm Selection Guidance}

FFT-modal is appropriate when uniform sampling and phase coherence are maintained, such as in conventional chamber measurements. Phaseless GS is suitable when phase-measurement hardware is unavailable but uniform grids are feasible. ASPIRE is recommended for drone-based platforms where positioning jitter, scan truncation, or irregular sampling demand a robust inverse-source formulation with explicit coupling resolution. Consequently, when reliable phase data is available, the general performance hierarchy is ASPIRE $\gg$ FFT-Modal $>$ Phaseless GS. Conversely, when phase information is unavailable or degraded, FFT-modal and Phaseless GS switch order, making phase retrieval the preferable baseline.

\section{Conclusion}

This paper compared three NF-FF transformation approaches under drone-induced measurement errors at C-band and X-band. ASPIRE, a cascaded randomized SVD $\ell_1$-regularized inverse-source solver with RWG basis functions and explicit mutual coupling, achieves $\leq 0.3^\circ$ beamwidth reconstruction error across both frequency bands, outperforming FFT-modal and phaseless alternatives in sidelobe fidelity and main-beam accuracy. MLFMM acceleration with Numba JIT compilation provides up to $15\times$ memory reduction, enabling scaling to $10^5$+ unknowns on commodity hardware. Future work includes incorporating controlled ablation studies to experimentally decouple the compounded penalties of positioning jitter and scan-area truncation, and extension to higher frequency bands.


\end{document}